\documentclass{lhep}       

\received{xx January 2018}
\published{xx March 2018}

\def\be{\begin{equation}}
\def\ee{\end{equation}}
\def\bea{\begin{eqnarray}}
\def\eea{\end{eqnarray}}

\newcommand{\T}[1]{\tilde{#1}}

\newcommand{\eval}[1]{\big<#1\big>}

\begin{document}

\title{Fierz identities at the one-loop level}

\author{Jason Aebischer}
\address{Physik-Institut, Universit\"at Z\"urich, CH-8057 Z\"urich, Switzerland}

\begin{abstract}
One-loop Fierz identities are discussed, together with general basis transformations in Effective Field theories at the tree- and one-loop level. To this end, the notion of one-loop shifts is introduced, together with several examples that illustrate the virtues of this method. 

\end{abstract}

\maketitle


\section{Introduction}
Higher-order loop calculations often involve the notion of Evanescent Operators (EVs) \cite{Buras:1989xd,Buras:2020xsm,Dugan:1990df,Herrlich:1994kh}, that are introduced in $d=4-2\epsilon$ dimensions in order to compensate for the breaking of four-dimensional identities and to express results in terms of physical operators. In the case where EVs are related to Fierz identities \cite{Fierz:1939zz}, their effects can be interpreted as one-loop shifts to the four-dimensional Fierz identities. Such shifts were discussed for the first time in \cite{Aebischer:2022tvz}, in the context of $\Delta F=1$ transitions and were used in several phenomenological analyses \cite{Aebischer:2021raf,Aebischer:2021hws,Aebischer:2020dsw,Aebischer:2018acj}. A systematic treatment of one-loop Fierz identities was first presented in \cite{Aebischer:2022aze}, where all one-loop QCD and QED shifts to four-fermi operators were considered. The discussion was generalized in \cite{Aebischer:2022rxf} to include effects from dipole operators. The work in \cite{Aebischer:2022aze,Aebischer:2022rxf} therefore concludes the discussion of one-loop Fierz relations in the Weak Effective Theory (WET) \cite{Aebischer:2017gaw} below the electroweak (EW) scale. Moreover, recent progress was made concerning various EV bases: The complete EV basis in the `t Hooft-Veltmann (HV) scheme for the WET was recently derived in \cite{Naterop:2023dek}. Above the EW scale in the Standard Model Effective Field Theory (SMEFT) the complete EV basis was presented in \cite{Fuentes-Martin:2022vvu}.

A typical application of one-loop Fierz identities are one-loop basis changes: In Effective Field Theories (EFTs) different computations are often performed in different operator bases. Often, the matching computation is performed in a basis that differs from the one that is used to compute the Anomalous Dimension Matrix (ADM) of the EFT. A common example is the matching of Leptoquarks (LQs) onto the Standard Model (SM), for which the matching is performed in the LQ basis, whereas the running is known in the SM basis \cite{Aebischer:2018acj}. At the one-loop level, in order to combine the two results a one-loop basis change has to be performed between the two bases.

In the following we will elaborate on the one-loop Fierz identities and their use in one-loop basis changes. Several examples from the literature will be discussed in order to illustrate the approach.

\section{One-loop Fierz identities}
Fierz identities relate four-dimensional Dirac structures to each other. A given four-fermi operator $\tilde Q$ is related to its Fierz-transformed version $Q$ in the following way:
\begin{equation}\label{eq:treeF}
	\tilde Q =\mathcal{F}Q\,,
\end{equation}
where $\mathcal{F}$ denotes the tree-level Fierz transformation between them. However, when loop effects are considered, spacetime is promoted from four to general $d=4-2\epsilon$ dimensions, in order to regularize the divergent integrals. In general $d$ dimensions the Fierz relation \eqref{eq:treeF} does not hold anymore, but has to be generalized to the $d$-dimensional case. In the traditional way this is achieved by introducing an evanescent operator $E$, such that the generalized Fierz identity reads:

\begin{equation}\label{eq:1LF}
	\tilde Q =\mathcal{F}Q+E\,,
\end{equation}

The newly introduced EV $E$ is simply defined via eq.~\eqref{eq:1LF}, i.e. as the difference between $\tilde Q$ and its Fierz-conjugate $\mathcal{F}Q$, and consequently vanishes in the limit $d\to 4$. The effect of this newly introduced EV can however be interpreted as a one-loop correction to the four-dimensional Fierz relation in eq.~\eqref{eq:treeF}. Using this method, instead of eq.~\eqref{eq:1LF}, the tree-level Fierz identity is generalized in the following way:

\begin{equation}\label{eq:1LFc}
	\tilde Q=\mathcal{F}Q +\sum_{b}\sum_i\frac{\alpha_b}{4\pi} a^{(b)}_i Q_i\,,
\end{equation}
where the index $i$ runs over the physical basis of the theory and $b$ labels different coupling constants. The scheme-dependent factors $a^{(b)}_i$ coincide with the contributions from evanescent operators. Notice, that in eq.~\eqref{eq:1LFc} no EV had to be introduced and hence no enlargement of the basis is needed. The effects of the EVs is encoded in the $a^{(b)}_i$ factors, which only have to be computed once for a given scheme. They are obtained by computing one-loop diagrams of the operator and its Fierz-conjugate and expressing the difference between the two in terms of the physical basis. In this calculation only the divergent parts of the loop integrals are needed, since the finite terms are equal in both bases and drop out in the difference.

The factors $a^{(b)}_i$ were computed for the case of one-loop QCD and QED corrections in \cite{Aebischer:2022aze}. The computation was performed for all possible four-quark, four-lepton and semileptonic operators, including scalar, vector and tensor Dirac structures. The used basis is given by
\begin{equation}\label{eq:basis}
	\{P_{A}\otimes P_{B},\,\, \gamma_\mu P_{A}\otimes \gamma^\mu P_{B},\,\, \sigma_{\mu\nu} P_{A}\otimes \sigma^{\mu\nu} P_{B}\}\,,
\end{equation} 
with the projection operators $P_{A,B}$ with $A,B=L,R$. In \cite{Aebischer:2022aze} several tables for all of the operators with the corresponding QCD and QED shifts are provided. The one-loop shifts are computed for the Fierzed versions of the basis operators in eq.~\eqref{eq:basis}, which are themselves four-fermion operators. In \cite{Aebischer:2022rxf} the computation was generalized to the case where also dipole operators contribute to the one-loop shifts, i.e. where an $i$-index in eq.~\eqref{eq:1LFc} denotes one of the dipole operators

\begin{align}
	D_{q_1q_2 G}^B &= \frac{1}{g_s}m_q (\overline q_1 \sigma^{\mu\nu}P_B T^A q_2)G^A_{\mu\nu}\,,\\
	D_{f_1f_2 \gamma}^B &= \frac{1}{e}m_f (\overline f_1 \sigma^{\mu\nu}P_B  f_2)F_{\mu\nu}\,,
  \end{align}
  with $f\in\{q,\ell\}$. Such contributions arise when mass effects are considered in one-loop penguin diagrams.

The results obtained in \cite{Aebischer:2022aze} are given in the $\overline{\rm MS}$-scheme and in the generalized BMU \cite{Buras:2000if} scheme. The one-loop shifts from dipole operators in \cite{Aebischer:2022rxf} were presented in the $\overline{\rm MS}$-scheme. Apart from being scheme-dependent, the results in \cite{Aebischer:2022aze,Aebischer:2022rxf} are general and can be used for arbitrary operator bases that might differ from the one in eq.~\eqref{eq:basis}. The one-loop shifts in the novel basis are simply obtained by performing a tree-level basis change to the bases and the shifts derived in \cite{Aebischer:2022aze,Aebischer:2022rxf}.

\section{One-loop basis change}
In the traditional way, basis changes at higher-orders are performed by relating physical as well as evanescent operators from two bases \cite{Chetyrkin:1997gb,Gorbahn:2004my}. For this purpose, various renormalization constants like the finite renormalization constant $Z_{EQ}$ needs to be computed, in order to accommodate for the scheme-dependent contributions from the EVs mixing into physical operators.

When using the one-loop shift method however, only physical operators need to be related to each other. Furthermore, using this method it is possible to change simultaneously the operator basis as well as the renormalization scheme. It was shown in \cite{Aebischer:2023djt} that in the framework of one-loop Fierz identities general one-loop basis changes can be parametrized in the following way:
\begin{equation}
	\vec{\T{Q}}_{\T{\Sigma};\T{S}} = \Big(\mathcal{F} + \Delta\Big)\vec{Q}_{\Sigma;S}\,,
\end{equation}
where the operator basis $\vec{\T{Q}}$ is associated to a scheme $\T{S}$ specifying the treatment of $\gamma_5$, as well as scheme-dependant contributions from EVs, parametrized by $\T{\Sigma}$, and similar expressions for the untilded case. Furthermore, as in eq.~\eqref{eq:treeF} $\mathcal{F}$ denotes the tree-level basis change and $\Delta$ stands for the shift. It is given at the one-loop level by the following expression:

\begin{equation}\label{eq:masterformula}
	\Delta\eval{\vec{Q}}^{(0)} = P_{Q;\Sigma}\Big(\eval{\vec{\T{Q}}}_{\T{\Sigma};\T{S}}^{(1)}\Big)
			- \Big(\mathcal{F} + \epsilon \Sigma\Big)\eval{\vec{Q}}^{(1)}_{\Sigma;S}\,,
\end{equation}
where $\eval{...}^{(0)}$ and $\eval{...}^{(1)}$ denote bare tree- and one-loop matrix elements, respectively. Furthermore, we have introduced the notation
\begin{equation}
	P_{Q;\Sigma}\,\mathcal{M}\Big(\eval{\vec{\T{Q}}}^{(0)}\Big) = \mathcal{M}\Big(\big(\mathcal{F} + \epsilon \Sigma\big)\eval{\vec{Q}}^{(0)}\Big)\,,
\end{equation}
which for a given matrix element $\mathcal{M}$ denotes the replacement $\eval{\vec{\T{Q}}}^{(0)}=\big(\mathcal{F} + \epsilon \Sigma\big)\eval{\vec{Q}}^{(0)}$ at the given loop order.
For further details of the method and a general proof of eq.~\eqref{eq:masterformula} we refer to \cite{Aebischer:2023djt}.

\section{Examples}

\subsection{Traces involving $\gamma_5$}
One example where one-loop basis changes become important is the case where traces involving the Dirac matrix $\gamma_5$ occur in the calculation. In the NDR renormalization scheme such traces are ill-defined and therefore lead to inconsistent results. Therefore, in order to avoid such expressions a basis change can be performed to a basis where such traces are absent. As an example we consider the tensor operator

\begin{equation}
	O^{T,RR}_{\substack{eu\\2222}} = (\overline \mu \sigma_{\mu\nu} P_R \mu)(\overline c \sigma^{\mu\nu} P_R c)\,,
\label{eq:badtensor}
\end{equation}
and its one-loop contribution to the muon anomalous magnetic moment $a_\mu=(g-2)_\mu$. When closing the charm-loop and attaching a photon this operator generates the electric dipole operator 

\begin{equation}
	D_{\mu\mu \gamma}^R =  (\overline \mu \sigma^{\mu\nu}P_R  \mu)F_{\mu\nu}\,.
\end{equation}
The one-loop contribution however contains the problematic trace 
\begin{equation}
	Tr[\gamma_\mu\gamma_\nu\gamma_\rho\gamma_\sigma\gamma_5]\,,
\end{equation}
which can not be evaluated in NDR. In order to be able to use NDR, the Fierzed operator 

\begin{equation}
	\tilde O^{T,RR}_{\substack{eu\\2222}} = (\overline \mu \sigma_{\mu\nu} P_R c)(\overline c \sigma^{\mu\nu} P_R \mu)\,,
\label{eq:badtensor}
\end{equation}
can be used, which only generates open-penguin structures which can easily be evaluated using NDR. To obtain the final result in terms of the initial operator $O^{T,RR}_{\substack{eu\\2222}}$ a one-loop basis change has to be considered. Combining the one-loop matrix element of $\tilde O^{T,RR}_{\substack{eu\\2222}}$ with the one-loop shift in \cite{Aebischer:2022rxf} one finds the following contribution to the anomalous magnetic moment of the muon:

\begin{equation}
	a^{2\ell 2q}_\mu = -m_\mu m_c\frac{N_c Q_c}{\pi^2 Q_\mu}\log\Big(\frac{\mu^2}{m_c^2}\Big)\Re \left[L^{T,RR}_{\substack{eu\\\mu\mu cc}}(\mu)\right]\,,
\end{equation}
which agrees with the result in \cite{Aebischer:2021uvt}.

\subsection{Different renormalization schemes} 
In order to avoid complicated renormalization schemes it is often more convenient to do the calculation in a simpler renormalization scheme and subsequently perform a scheme change to the target scheme. As an example we consider the two-loop QCD ADM of the semi-leptonic operators 

\begin{align}\label{eq:lar_defs}
	\mathcal{O}_1^{ij} &= \big(\bar{\psi}_i\psi_i\big)\big(\bar{\psi}_j\,i \gamma_5\,\psi_j\big), \quad
	\mathcal{O}_2 = \frac{1}{2}\epsilon^{\mu \nu \rho \sigma}\big(\bar{e}\sigma_{\mu \nu}e\big)\big(\bar{b}\sigma_{\rho\sigma}b\big), \notag\\
	\mathcal{O}_3 &= \frac{Q_e m_b}{2 e}\big(\bar{e}\sigma^{\mu \nu}e\big)\tilde{F}_{\mu \nu}\,,
\end{align}
with $\psi_{i,j} = e$ or $b$. In \cite{Brod:2023wsh} this calculation was performed in the Larin scheme due to the appearance of trances with $\gamma_5$. It is however possible to compute the ADM in the much simpler NDR-$\overline{\rm MS}$ scheme, using the operator basis
\begin{equation}\label{eq:PHYSbasis}
	\begin{split}
		\T{\mathcal{O}}_s &= \frac{1}{2}\Big[\big(\bar{b}i\gamma_5 e\big)\big(\bar{e} b\big) 
			+ \big(\bar{b} e\big)\big(\bar{e}i\gamma_5 b\big)\Big]\,,\quad \T{\mathcal{O}}_3 = \frac{Q_e m_b}{2e}\big(\bar{e}i\sigma_{\mu \nu}\gamma_5 e\big) F^{\mu \nu} \\
		\T{\mathcal{O}}_v &= \frac{1}{2}\Big[\big(\bar{b}i\gamma^\mu \gamma_5 e\big)\big(\bar{e}\gamma_\mu b\big) 
			- \big(\bar{b}\gamma^\mu e\big)\big(\bar{e}i\gamma_\mu \gamma_5 b\big)\Big]\,, \\[0.5em]
		\T{\mathcal{O}}_t &= \frac{1}{2}\Big[\big(\bar{b}i\sigma_{\mu \nu}\gamma_5 e\big)\big(\bar{e}\sigma^{\mu \nu} b\big) 
			+ \big(\bar{b}\sigma_{\mu \nu} e\big)\big(\bar{e}i\sigma^{\mu \nu}\gamma_5 b\big)\Big]\,, 
	\end{split}
\end{equation}
The one-loop shift between the two bases and schemes is given in eq.~\eqref{eq:masterformula} and reads for the case at hand:

\begin{equation}\label{eq:Delta}
	\Delta = \Delta^{\text{SI}} + a_s\Delta^{s} + a_v \Delta^{v} + a_t \Delta^{t}\,,
\end{equation}
where $\Delta^{\text{SI}}$ denotes the scheme-independent shift and $a_i$ are scheme-dependent constants for the respective Dirac structures. The individual shifts are given by \cite{Aebischer:2023djt}

\begin{equation}
	\begin{split}
		\Delta^{\text{SI}} =& \begin{pmatrix}
			\frac{3}{2} & -\frac{7}{6} & \frac{1}{12} & 0 \\[0.5em]
			-\frac{28}{3} & -\frac{4}{3} & 0 & 0 \\[0.5em]
			\frac{110}{3} & \frac{14}{3} & -\frac{13}{3} & 0 \\[0.5em]
			0 & 0 & 0 & 0
		\end{pmatrix},\,\, \Delta^{v} = \begin{pmatrix}
			0 & 0 & 0 & 0 \\[0.5em]
			-\frac{1}{3} & \frac{1}{3} & 0 & 0 \\[0.5em]
			0 & 0 & 0 & 0 \\[0.5em]
			0 & 0 & 0 & 0
		\end{pmatrix}\,, \\[1.0em]
		\Delta^{s} =& \begin{pmatrix}
			0 & 0 & 0 & 0 \\[0.5em]
			0 & 0 & 0 & 0 \\[0.5em]
			-\frac{1}{12} & -\frac{1}{12} & -\frac{1}{24} & 0 \\[0.5em]
			0 & 0 & 0 & 0
		\end{pmatrix},\,\, \Delta^{t} = \begin{pmatrix}
			0 & 0 & 0 & 0 \\[0.5em]
			0 & 0 & 0 & 0 \\[0.5em]
			1 & 1 & -\frac{1}{6} & 0 \\[0.5em]
			0 & 0 & 0 & 0
		\end{pmatrix}\,.
	 \end{split}
\end{equation}

Combining the two-loop ADM in the NDR scheme together with the shift in eq.~\eqref{eq:Delta} leads to the results in \cite{Brod:2023wsh}. For more details on the calculation and renormalization procedure we refer to \cite{Aebischer:2023djt}.

\section{Summary}

One-loop Fierz identities have been discussed for the complete operator basis below the EW scale in \cite{Aebischer:2022aze,Aebischer:2022rxf}. The results can be applied to arbitrary operator bases by changing the target basis at {\it tree-level} to the bases in \cite{Aebischer:2022aze,Aebischer:2022rxf}. Such one-loop shifts are useful when considering one-loop basis changes as well as scheme changes and it has been shown in \cite{Aebischer:2023djt} that both transformations can be performed simultaneously in this framework.
An automation of the findings discussed in this letter would be of great use for future EFT calculations. The authors plan to include the findings in the package \texttt{abc\_eft} \cite{Proceedings:2019rnh}, the matchrunner \texttt{wilson} \cite{Aebischer:2018bkb} as well as the basis change package \texttt{WCxf} \cite{Aebischer:2017ugx}.

\section*{Acknowledgements}
The work of J.\ A. is supported from  the  European  Research  Council  (ERC)  under the European Union's Horizon 2020 research and innovation programme under grant agreement 833280 (FLAY), and from the Swiss National Science Foundation (SNF) under contract 200020-204428.

\bibliographystyle{JHEP}

\bibliography{refs}

\end{document}